\documentclass[floatfix,aps,prl,showpacs,amsmath,nofootinbib,preprintnumbers,twocolumn]{revtex4}

\begin{document}

\title{A Weak Gravity Conjecture for Scalar Field Theories}

\author{Miao Li$^{1,2,3}$}\email{mli@itp.ac.cn}
\author{Wei Song$^{1,3}$}\email{wsong@itp.ac.cn}
\author{Yushu Song$^{1,3}$}\email{yssong@itp.ac.cn}
\author{Tower Wang$^{1,3}$}\email{wangtao218@itp.ac.cn}

\affiliation{$^1$ Institute of Theoretical Physics, Academia Sinica, Beijing 100080, China\\
$^2$ Interdisciplinary Center of Theoretical Studies, Academia Sinica, Beijing 100080, China\\
$^3$~Interdisciplinary Center for Theoretical Study, University of Science and Technology of China,\\
Hefei, Anhui 230026, China}

\date{\today}

\begin{abstract}
We put forward a conjecture for a class of scalar field theories
analogous to the recently proposed weak gravity conjecture
\cite{AMNV0601} for U(1) gauge field theories. Taking gravity into
account, we find an upper bound on the gravity interaction strength,
expressed in terms of scalar coupling parameters. This conjecture is
supported by some two-dimensional models and noncommutative field
theories.
\end{abstract}

\maketitle

%%%%%%%%%%%%%%%%%%%%%%%%%%%%%%%%%%%%%%%%
%%%%%%%%%%%%%%%%%%%%%%%%%%%%%%%%%%%%%%%%
Scalars, such as Higgs and inflaton, play significant roles in both
particle physics and cosmology. Recently, Arkani-Hamed et.al.
conjectured an upper bound on the strength of gravity relative to
gauge forces in quantum gravity \cite{AMNV0601}. Their conjecture
enriches the criteria of consistent effective field theory proposed
in \cite{D0507,V0509}. These criteria help to constrain the string
landscape \cite{S0302} in the vast vacua of string theory. After a
careful investigation, we find that a similar bound exists in a
class of scalar field theories. In this class of theories at least,
``gravity is the weakest'' even in the appearance of scalars.

We will study scalar field theories with soliton solutions. we will
focus our attention on a special class of such theories. In these
theories, the coefficient of the mass term is $\pm\frac{1}{2}\mu^2$
with $\mu^2>0$, and there are higher order terms controlled by a
coupling constant $\lambda$ in addition to $\mu$. So the theories
are described by only two parameters: the mass parameter $\mu$ and
the coupling constant $\lambda$. When the scalar is coupled to
gravity, we find an interesting constraint on $\mu$ and $\lambda$.

The idea is to study a scalar system coupled to two-dimensional
dilaton gravity, which is assumed to be dimensional reduction of
some consistent quantum gravity system in higher dimensions. In two
dimensions, the system contains solitons in the weak gravity limit
(plus some other massive particles in a sector not considered here).
These solitons appear as asymptotic scattering states so there is a
well-defined S-matrix. As one increases the gravity coupling, these
solitons disappear, thus there are no more massive particles and the
S-matrix ceases to exist. We assume that in this case the quantum
system becomes inconsistent, because without the S-matrix, only
correlation functions exist which are not observables in a gravity
theory. We therefore conjecture that the quantum theory exists only
in the weak gravity region. In fact, in some exact S-matrices, such
as the Sine-Gordon model \cite{KF75,Z77} and the chiral field model
\cite{W84}, only solitons appear in spectra. It is a matter subject
to debate that whether this is also true in higher dimensions, but
we feel that a consistent theory in higher dimensions should result
in a consistent theory upon dimensional reduction, thus we lift this
conjecture to higher dimensions, in particular, to four dimensions.

When the scalar is coupled to two-dimensional dilaton gravity, the
action is generally taken to be
\begin{eqnarray}\label{2d act1}
\nonumber S&=&\frac{1}{2\pi}\int d^{2}x\sqrt{-\gamma}e^{-2\phi}[\frac{1}{8G}\left(R+4({\nabla\phi})^2\right)\\
&&-\frac{1}{2}({\nabla\varphi})^2\mp\frac{1}{2}\mu^2\varphi^2-V_{c}(\varphi,\lambda,\mu)]
\end{eqnarray}

In this notation, the potential of scalar $\varphi$ is
$V=V_{c}\pm\frac{1}{2}\mu^2\varphi^2$. In most cases, after a
redefinition of $\lambda$, the higher order coupling terms $V_{c}$
can be rewritten as
\begin{equation}\label{lambda}
V_{c}(\varphi,\lambda,\mu)=\frac{\mu^2}{g^{2\alpha}}{V'}_{c}(g^{\alpha}\varphi)
\end{equation}
in which $g=\frac{\sqrt{\lambda}}{\mu}$, and the value of $\alpha$
depends on the form of $V_{c}$. As we will show later, for a given
$V_{c}$, the value of $\alpha$ can be worked out explicitly.

We rescale the coordinates and the scalar to make them dimensionless
\begin{equation}
t'=\mu t,~~~~x'=\mu x,~~~~\varphi'=g^{\alpha}\varphi
\end{equation}
the action (\ref{2d act1}) becomes
\begin{eqnarray}\label{2d act2}
\nonumber S&=&\frac{1}{2\pi}\int d^{2}x'\sqrt{-\gamma}e^{-2\phi}[\frac{1}{8G}\left(R'+4(\nabla'\phi)^2\right)\\
&&+\frac{1}{g^{2\alpha}}\left(-\frac{1}{2}({\nabla'\varphi'})^2\mp\frac{1}{2}\varphi'^2-{V'}_{c}(\varphi')\right)]
\end{eqnarray}

We stress that rescaling $\lambda$ in (\ref{lambda}) by a numeric
factor will rescale the expression of ${V'}_{c}$ also by a numeric
factor. To remove this uncertainty, when writing (\ref{lambda}), one
should define $\lambda$ appropriately to get a reasonable ${V'}_{c}$
whose numeric coefficients are of order 1.

From (\ref{2d act2}), it is clear that the gravitational strength
relative to the strength of scalar interaction is controlled by
$\frac{G}{g^{2\alpha}}$. If $g^{2\alpha}\gg G$, the effect of
gravity is just a small correction to the original scalar field
theory, so one can treat the weak gravity perturbatively, expanding
in terms of $\frac{\sqrt{G}}{g^{\alpha}}$. On the other hand, if
$g^{2\alpha}\ll G$, the correction would be large enough to destroy
the original solutions and make the theory inconsistent. Hence
quantum gravity provides a criterion of consistent scalar field
theories, parametrically
\begin{equation}
g^{2\alpha}\gtrsim G
\end{equation}
This constraint can be lifted to higher dimensions:
\begin{equation}\label{conject1}
(\frac{\lambda}{\mu^2})^{\alpha}\gtrsim G
\end{equation}
In $D$-dimensional spacetime, the Newton constant $G$ has the
dimensions of $mass^{2-D}$ while the coupling constant $\lambda$ has
the dimensions of $mass^{2-\frac{D-2}{\alpha}}$.

In the special case
\begin{equation}
V=-\frac{1}{2}\mu^2\varphi^2+\frac{1}{n}\lambda\varphi^{n}+(\frac{1}{2}-\frac{1}{n})\mu^2(\frac{\mu^2}{\lambda})^\frac{2}{n-2}
\end{equation}
with an even number $n>2$, we have $\alpha=\frac{2}{n-2}$,
therefore the relation (\ref{conject1}) becomes
\begin{equation}
(\frac{\lambda}{\mu^2})^\frac{2}{n-2}\gtrsim G
\end{equation}
In four dimensions, setting $n=4$ leads to the
constraint $\frac{\mu}{\sqrt{\lambda}}\lesssim M_{Pl}$.

In the examples studied below, when gravity decouples, solitons are in the
spectrum. Disappearance solitons in the strong gravity singals
breakdown of unitarity: when we tune up gravity interaction strength gradually, the rank of S-matrix will
have a jump at some point in the parameter space. Indeed, we do not
even know if the S-matrix exists in the strong gravity region. So
for the parameter region violating (\ref{conject1}), there is no
consistent quantum gravity theory. We suspect that the same kind of relation
(\ref{conject1}) remains true in a larger class of theories without
solitons, although we do not have evidence.

A kink is the most familiar soliton in two-dimensional spacetime.  The
action of the coupled system reads
\begin{equation}\label{kink act}
S=\frac{1}{2\pi}\int d^{2}x\sqrt{-\gamma}e^{-2\phi}[R+4({\nabla\phi})^2-\frac{1}{2}({\nabla\varphi})^2-V]
\end{equation}
the equations of motion are
\begin{eqnarray}\label{kink eom}
\nonumber R+4\nabla^2\phi-4(\nabla\phi)^2-\frac{1}{2}(\nabla\varphi)^2-V&=&0 \\
\nonumber 2\gamma^{ab}\nabla_a\varphi\nabla_b\phi-\nabla^2\varphi+\frac{dV}{d\varphi}&=&0\\
\nonumber -\frac{1}{2}\nabla_a\varphi\nabla_b\varphi+2\nabla_a\nabla_b\phi&&\\
+\gamma_{ab}[2(\nabla\phi)^2-2\nabla^2\phi+\frac{1}{4}(\nabla\varphi)^2+\frac{1}{2}V]&=&0
\end{eqnarray}
with $a,b=0,1$ in two dimensions.

We are interested in static solutions, in which the dilaton
$\phi$, the scalar field $\varphi$ and the metric $\gamma_{ab}$ are
independent of $t$.

For any two-dimensional static metric
\begin{eqnarray}\label{staticg}
\nonumber ds^2&=&\gamma_{00}dt^2+2\gamma_{01}dtdx+\gamma_{11}dx^2\\
&=&-\gamma_{00}[-(dt+\frac{\gamma_{01}}{\gamma_{00}}dx)^2+\frac{\gamma_{01}^2-\gamma_{00}\gamma_{11}}{\gamma_{00}^2}dx^2]
\end{eqnarray}
By redefining coordinates
\begin{equation}\label{redefc}
d\tilde{t}=dt+\frac{\gamma_{01}}{\gamma_{00}}dx,~~~~d\tilde{x}=\sqrt{\frac{\gamma_{01}^2-\gamma_{00}\gamma_{11}}{\gamma_{00}^2}}dx
\end{equation}
one can always switch to the conformal gauge
$\tilde{\gamma}_{ab}=e^{2w}\eta_{ab}$ while keeping the solutions
static. Henceforce, we will work in conformal gauge and omit the
tilde for brevity.

The equations of motion (\ref{kink eom}) now reduce to
\begin{eqnarray}
2(\frac{d\phi}{dx})^2-2\frac{d^2\phi}{dx^2}+\frac{1}{4}(\frac{d\varphi}{dx})^2+\frac{1}{2}e^{2w}V&=&-\frac{d^2w}{dx^2}\label{eom1}\\
2(\frac{d\phi}{dx})^2-2\frac{d^2\phi}{dx^2}+\frac{1}{4}(\frac{d\varphi}{dx})^2+\frac{1}{2}e^{2w}V&=&-2\frac{dw}{dx}\frac{d\phi}{dx}\label{eom2}\\
\nonumber 2(\frac{d\phi}{dx})^2-2\frac{d^2\phi}{dx^2}+\frac{1}{4}(\frac{d\varphi}{dx})^2+\frac{1}{2}e^{2w}V&&\\
=2\frac{dw}{dx}\frac{d\phi}{dx}+\frac{1}{2}(\frac{d\varphi}{dx})^2&-&2\frac{d^2\phi}{dx^2}\label{eom3}\\
2\frac{d\varphi}{dx}\frac{d\phi}{dx}-\frac{d^2\varphi}{dx^2}+e^{2w}\frac{dV}{d\varphi}&=&0\label{eom4}
\end{eqnarray}
We have written them in forms that the left hand sides of
(\ref{eom1}), (\ref{eom2}) and (\ref{eom3}) are the same. Equating
the right hand sides of (\ref{eom1}) and (\ref{eom2}), we find
a simple relation
\begin{equation}\label{simp eq1}
\frac{dw}{dx}=Ce^{2\phi}
\end{equation}
Combining it with (\ref{eom2}) and (\ref{eom3}), it follows that
\begin{equation}\label{simp eq2}
4Ce^{2\phi}\frac{d\phi}{dx}=2\frac{d^2\phi}{dx^2}-\frac{1}{2}(\frac{d\varphi}{dx})^2
\end{equation}
Using (\ref{eom2}) and (\ref{eom3}) to eliminate $\frac{dw}{dx}$, we
obtain
\begin{equation}
4(\frac{d\phi}{dx})^2-2\frac{d^2\phi}{dx^2}+e^{2w}V=0
\end{equation}
Combined with (\ref{eom4}) to eliminate $e^{2w}$, it gives
\begin{equation}\label{simp eq3}
2\frac{dV}{d\varphi}[\frac{d^2\phi}{dx^2}-2(\frac{d\phi}{dx})^2]=V(\frac{d^2\varphi}{dx^2}-2\frac{d\varphi}{dx}\frac{d\phi}{dx})
\end{equation}

We will restrict our attention to a kink potential $V(\varphi)$ with
$\alpha=1$ in (\ref{lambda}). Based on the work of 't
Hooft \cite{Hooft}, we assume
\begin{equation}\label{lead varphi}
\varphi=\frac{1}{g}f_{-1}(x)+\mathrm{subleading~terms}
\end{equation}
where we have used the notation of $g=\frac{\sqrt{\lambda}}{\mu}$,
while $f_{-1}(x)$ is some function independent of $g$.

Remember that in the notation of (\ref{2d act2}), the potential and
its derivative with respect to $\varphi$ can be written in the form
\begin{eqnarray}
\nonumber V(\varphi,\lambda,\mu)&=&\frac{\mu^2}{g^2}V'(g\varphi)=\frac{\mu^2}{g^2}V'(\varphi')\\
\frac{dV}{d\varphi}&=&\frac{\mu^2}{g^2}\frac{dV'}{d\varphi'}\frac{d\varphi'}{d\varphi}=\frac{\mu^2}{g}\frac{dV'}{d\varphi'}
\end{eqnarray}
In the leading order $\varphi'\sim f_{-1}$, therefore
\begin{eqnarray}\label{lead v}
\nonumber V&=&\frac{\mu^2}{g^2}V'(f_{-1})+\mathrm{subleading~terms}\\
\frac{dV}{d\varphi}&=&\frac{\mu^2}{g}\frac{dV'}{df_{-1}}+\mathrm{subleading~terms}
\end{eqnarray}
Substitute  the leading order terms of (\ref{lead varphi}) and
(\ref{lead v}) into (\ref{simp eq3}), We find
\begin{equation}
2\frac{dV'}{df_{-1}}[\frac{d^2\phi}{dx^2}-2(\frac{d\phi}{dx})^2]=\frac{1}{g^2}V'(f_{-1})(\frac{d^2f_{-1}}{dx^2}-2\frac{df_{-1}}{dx}\frac{d\phi}{dx})
\end{equation}
In order to be consistent with this equation, $\phi$ has to take the
form
\begin{equation}\label{lead phi}
\phi=\frac{1}{g^2}\Phi_{-2}+\mathrm{subleading~terms}
\end{equation}

In the strong gravity limit $g\ll1$, the value of $\phi$ in (\ref{lead
phi}) will generate a divergent value of $w$ in (\ref{simp eq1}),
unless $C=0$. So we set $C=0$ from now on in order to avoid the
essential singularity.

We shall use the perturbative expansion to study
equations (\ref{simp eq2}) and (\ref{simp eq3}) with $C=0$ in the weak
gravity limit and  thestrong gravity limit respectively.

In the weak gravity region, we assume $g\gg1$ and expand
$\phi$ and $\varphi$ in $\frac{1}{g}$
\begin{eqnarray}\label{weak phi}
\nonumber \varphi &=& \frac{1}{g}f_{-1}+\frac{1}{g^2}f_{-2}+... \\
\phi &=& \frac{1}{g^2}\Phi_{-2}+\frac{1}{g^3}\Phi_{-3}+...
\end{eqnarray}
Substituting (\ref{weak phi}) into (\ref{simp eq2}) and (\ref{simp
eq3}), we have the following lowest order equations
\begin{eqnarray}\label{weak eom}
\nonumber \frac{d^2\Phi_{-2}}{dx^2} &=& \frac{1}{4}(\frac{df_{-1}}{dx})^2\\
\frac{d^2\Phi_{-2}}{dx^2}&=&\frac{1}{2}\frac{V}{\frac{dV}{df_{-1}}}\frac{d^2f_{-1}}{dx^2}
\end{eqnarray}
So we have
\begin{equation}\label{weak scl}
\frac{V}{\frac{dV}{df_{-1}}}\frac{d^2f_{-1}}{dx^2}=\frac{1}{2}(\frac{df_{-1}}{dx})^2
\end{equation}
It is the same as the kink equation in \cite{Hooft} after a simple
calculation
\begin{equation}
\frac{d\varphi}{dx}=\sqrt{2V}
\end{equation}
We conclude that in weak gravity region the kink solution survives
when this scalar field is coupled to gravity.

The equation of motion  of the scalar field (\ref{weak scl}) reduce
to the equations of motion for scalar field theory in flat
spacetime. So if there are 2-soliton solutions in the pure scalar
field theory, there should be when the scalar field theory is
coupled to gravity. On the other hand, from (\ref{weak eom}) we see
that even there is no soliton, there may be linear dilaton, and it
does no harm to the one soliton soluton. With the soliton, the
second derivative of the dilaton becomes positive in some localized
region, and zero elsewhere. So asymptotically the soliton will
induce another linear dilaton. But the effect is only a shift of the
coefficient of the linear dilaton in pure gravity, and hence it will
do harm to two soliton solution. Thus, if S matrix exists in the
original pure scalar field theory, it will exist when the scalar
field theory is coupled to dilaton gravity.

In the strong gravity region, we take the limit $g\ll1$ and
expand $\phi$ and $\varphi$ in $g$
\begin{eqnarray}\label{str phi}
\nonumber \varphi &=& \frac{1}{g}f_{-1}+f_{0}+... \\
\phi &=& \frac{1}{g^2}\Phi_{-2}+\frac{1}{g}\Phi_{-1}+...
\end{eqnarray}
Plugging (\ref{str phi}) into (\ref{simp eq2}) and (\ref{simp eq3})
leads to the following lowest order equations
\begin{eqnarray}\label{str lowest}
\nonumber \frac{d^2\Phi_{-2}}{dx^2} &=& \frac{1}{4}(\frac{df_{-1}}{dx})^2\\
\frac{d\Phi_{-2}}{dx}&=&\frac{1}{2}\frac{V}{\frac{dV}{df_{-1}}}\frac{df_{-1}}{dx}
\end{eqnarray}
We will analyze the equations in two particular models, i.e. the
$\varphi^4$ model and the Sine-Gordon model.

\begin{enumerate}
\item $V(\varphi)=\frac{\lambda}{4}(\varphi^2-\frac{\mu^2}{\lambda})^2$\\
Substituting this potential into the equations (\ref{str lowest}),
we have the solution
\begin{eqnarray}\label{str kink}
\nonumber f_{-1} &=& e^{-8C_1x+C_2} \\
\Phi_{-2} &=& \frac{1}{16}e^{-16C_1x+2C_2}+C_1x+C_3
\end{eqnarray}
From (\ref{str kink}), it can be seen easily that $f_{-1}$ does not
satisfy the kink condition, so the kink solution breaks down in this case.\\
\item $V(\varphi)=\frac{\mu^4}{\lambda}[1-\cos(\frac{\sqrt{\lambda}}{\mu}\varphi)]$\\
Inserting this potential into the equations (\ref{str lowest}), we
have the solution
\begin{eqnarray}\label{str sine}
\nonumber f_{-1}&=& 2\arctan\sinh(D_1x+D_2)\\
\Phi_{-2}& =&\ln\cosh(D_1x+D_2)+D_3
\end{eqnarray}
From the potential in this case, we find that this solution connects
two maximum points rather than minimum, so there is no kink solution
in this case.
\end{enumerate}

In action (\ref{kink act}) we have set $8G=1$. By recovering $G$,
one finds that the expansion parameter is not $g$ but
$\frac{g}{\sqrt{G}}$. This accounts for the name ``weak/strong
gravity''.

The calculation above tells us that kink solutions appear in weak
gravity limit $\frac{g}{\sqrt{G}}\gg1$ but disappear in strong
gravity limit $\frac{g}{\sqrt{G}}\ll1$. In the absence of gravity,
the existence of soliton is dictated by unitarity. Naturally we
expect that unitarity will still require solitons when gravity is
switched on. Therefore, when $\frac{g}{\sqrt{G}}\ll1$ the breakdown
of kink solutions implies that we cannot take the strong gravity
limit smoothly. In other words, there must be a lower bound on the
value of $\frac{g}{\sqrt{G}}$, which is determined by the critical
point of the equations of motion. Parametrically, the critical value
is of order 1. This leads to
\begin{equation}
(\frac{\lambda}{\mu^2})\gtrsim G
\end{equation}
which agrees with the conjecture (\ref{conject1}). It would be
useful to work out the exact value of $\frac{g}{\sqrt{G}}$ at the
critical point.

All the discussions above are on the classical level. One
may wonder whether the conclusion is reliable on the quantum level.
To check this, we should consider some quantum corrections such as
higher derivative corrections. When we consider the original action
with the correction term $(\alpha'R)^n$, then the equations of
motion (\ref{eom1}), (\ref{eom2}) and (\ref{eom3}) will deserve the
correspondent correction terms, proportional to some positive power
of $R$. Note that our discussion below (\ref{redefc}) shows that the
we can always take the conformal gauge and set the field static in 2
dimensional spacetime. So we will always have
$R=-2e^{-2\omega}{d^2\omega\over dx^2}$. So fortunately these terms
will vanish under the classical solution (\ref{simp eq1}) with
$C=0$, e.g, ${d\omega\over dx}=0$. So we can conclude that our
results must be preserved at the quantum level.

Another piece of evidence comes from noncommutative solitons. In
noncommutative field theories, when the noncommutativity parameter
$\theta$ is sufficiently large (but finite), there are stable
soliton solutions in some three-dimensional scalar
theories \cite{GMS0106,DN0106}. Based on a cubic potential, in
\cite{HS0606}, Huang and She found a relation involving the
noncommutative parameter. In the following, we will briefly repeat
part of their work.

Consider a scalar field theory in three-dimensional noncommutative
spacetime, whose Euclidean action is
\begin{equation}\label{3d act}
S=\int d^3 x \sqrt{-\gamma}\left({1\over2}\gamma^{ij}\partial_{i}\varphi\partial_{j}\varphi+V(\varphi)\right).
\end{equation}
Written in terms of the canonically commuting noncommutative
coordinates
\begin{equation}\label{zzbar}
z={{x^1+ix^2}\over\sqrt{\theta}},~~\overline{z}={{x^1-ix^2}\over{\sqrt \theta}},
\end{equation}
the energy is
\begin{equation}\label{energy}
E=\int d^2 z\left({1\over2}(\partial\varphi)^2+\theta V(\varphi)\right).
\end{equation}
When $\theta V$ is large, the potential energy dominates and we can
find an approximate solitonic solution by solving the equation
${dV\over d\phi}=0$.

In our notations, the scalar potential is
$V(\varphi)=\frac{1}{2}\mu^2\varphi^2+\frac{1}{3}\lambda\varphi^{3}$.
The value of the potential at the stationary point
$\frac{dV}{d\varphi}=0$, namely
$\varphi_{cr}=-\frac{\mu^2}{\lambda}$, determines the soliton energy
completely \cite{DN0106}
\begin{equation}\label{mass}
E=2\pi\theta V(\varphi_{cr})=\frac{\pi}{3}\frac{\theta\mu^6}{\lambda^2}
\end{equation}
In three-dimensional spacetime, a massive pointlike soliton will
generate a deficit angle in the metric. The requirement that the
deficit angle be less than $2\pi$ implies
\begin{equation}\label{deficit}
8\pi GE<2\pi
\end{equation}
Observing (\ref{mass}) and (\ref{deficit}), we obtain
\begin{equation}
\frac{1}{G}\gtrsim\frac{\theta\mu^6}{\lambda^2}
\end{equation}

Actually, for a general potential (not necessarily of polynomial
form, see \cite{GMS0106}) with a noncommutative soliton whose energy
is positive
\begin{equation}\label{poten}
V=\frac{\mu^2}{g^{2\alpha}}{V'}_{c}(g^{\alpha}\varphi)\pm\frac{1}{2}\mu^2\varphi^2=\frac{\mu^2}{(\frac{\lambda}{\mu^2})^{\alpha}}[{V'}_{c}(\varphi')\pm\frac{1}{2}\varphi'^2]
\end{equation}
there is a similar relation
\begin{equation}\label{noncom}
\frac{1}{G}\gtrsim E\sim\frac{\theta\mu^2}{(\frac{\lambda}{\mu^2})^{\alpha}}
\end{equation}

On the other hand, the condition for the existence of soliton
solution in the noncommutative theory is
\begin{equation}\label{condi}\theta\geq\frac{1}{ \mu^2}.\end{equation}
Combine (\ref{noncom}) and (\ref{condi}), a constraint with the same
form of (\ref{conject1}) can be obtained. Using
(\ref{condi}) on the right hand side of (\ref{noncom}), we get
\begin{equation}\label{add}
\frac{1}{G}\gtrsim\frac{\theta\mu^2}{(\frac{\lambda}{\mu^2})^{\alpha}}\geq\frac{1}{(\frac{\lambda}{\mu^2})^{\alpha}},
\end{equation}
which is just the the form of (\ref{conject1}). This is a hint that
our conjecture (\ref{conject1}) is universal.

One may notice that in U(1) gauge field theories the weak gravity
conjecture has a sharp form \cite{AMNV0601}: there are always
particles with $M\lesssim Q$. However, it is difficult to similarly
sharpen our conjecture (\ref{conject1}). The particle charge $Q$ is
naturally defined for U(1) gauge field theories but not for scalar
field theories. Fortunately, it has been shown that in
\cite{AMNV0601} their conjecture can be expressed in different
forms, one of which is $T_{el}\lesssim \frac{g}{\sqrt{G}}$. This
form is similar to conjecture (\ref{conject1}) for scalar field
theories. Essentially, in both U(1) gauge field theories and scalar
field theories, the weak gravity conjecture is a statement about
effective field theories. So it can be used to identify the
swampland, a series of effective field theories which are consistent
semiclassically yet inconsistent in a quantum gravity theory.

We have so far been dealing with real scalar field theories. The
generalization to theories of complex scalars and scalar multiplets
is not straightforward. However, we have some hints. In \cite{LSW0601},
we have offered some evidence supporting the weak gravity conjecture
of gauge field theories. To avoid possible confusion, let us use $e$
to replace $g$ in \cite{LSW0601}. By noticing
$m_{W}^2=2e^2\frac{\mu^2}{\lambda}$, one can easily obtain the
relation (\ref{conject1}) for three-dimensional examples studied
in \cite{LSW0601}.

In \cite{HLS0603}, the weak gravity conjecture in \cite{AMNV0601}
has been generalized from flat spacetime to dS/AdS spacetime,
leading a bound on cosmological constant. It is also interesting to study
scalar theory in a dS background.

\textbf{Acknowledgments}. We would like to thank Qing-Guo Huang,
Jian-Huang She and Peng Zhang for useful discussions. This work was
supported by grants from CNSF.
%%%%%%%%%%%%%%%%%%%%%%%%%%%%%%%%%%%%%%%%
%%%%%%%%%%%%%%%%%%%%%%%%%%%%%%%%%%%%%%%%


\begin{thebibliography}{99}
\frenchspacing

\bibitem{S0302}
L. Susskind, ``The Anthropic Landscape of String Theory,''
hep-th/0302219.
%%CITATION =HEP-TH 0302219;%%

\bibitem{D0507}
M. R. Douglas,
``Is the number of string vacua finite?''
talk at the Strings 2005 Conference,\\
http://www.fields.utoronto.ca/audio/05-06/strings/douglas/

\bibitem{V0509}
C. Vafa,
``The String Landscape and the Swampland,''
hep-th/0509212.
%%CITATION =HEP-TH 0509212;%%

\bibitem{AMNV0601}
N. Arkani-Hamed, L. Motl, A. Nicolis and C. Vafa,
``The String Landscape, Black Holes and Gravity as the Weakest Force,''
hep-th/0601001.
%%CITATION =HEP-TH 0601001;%%

\bibitem{LSW0601}
M. Li, W. Song and T. Wang,
``Some Low Dimensional Evidence for the Weak Gravity Conjecture,''
JHEP {\bf 0603}, 094 (2006), hep-th/0601137.
%%CITATION = HEP-TH 0601137;%%

\bibitem{Hooft}
G. 't Hooft, ``Under the Spell of the Gauge Principle,'' World
Scientific.

\bibitem{GMS0106}
R. Gopakumar, S. Minwalla and A. Strominger ,
``Noncommutative Solitons,''
JHEP {\bf 0005}, 020 (2000), hep-th/0003160.
%%CITATION = HEP-TH 0003160;%%

\bibitem{DN0106}
M. R. Douglas and N. A. Nekrasov,
``Noncommutative Field Theory,''
Rev. Mod. Phys. {\bf 73}, 977 (2001), hep-th/0106048.
%%CITATION = HEP-TH 0106048;%%

\bibitem{HS0606}
Q. G. Huang and J. H. She,
``Weak Gravity Conjecture for Noncommutative Field Theory,"
JHEP {\bf 0612}, 014 (2006), hep-th/0611211.
%%CITATION = HEP-TH 0611211;%%

\bibitem{HLS0603}
Q. G. Huang, M. Li and W. Song,
``Bound on the U(1) gauge coupling in the asymptotically dS and AdS background,''
JHEP {\bf 0610}, 059 (2006), hep-th/0603127.
%%CITATION = HEP-TH 0603127;%%

\bibitem{KF75}
V. E. Korepin, L. D. Faddeev,
``Quantization Of Solitons,''
Teor. Mat. Fiz. {\bf 25}, 147 (1975).

\bibitem{Z77}
A.B. Zamolodchikov,
``Exact Two Particle S Matrix Of Quantum Sine-Gordon Solitons,''
Commun. Math. Phys. {\bf 55}, 183 (1977).

\bibitem{W84}
P. Wiegmann,
``Exact factorized S-matrix of the chiral field in two dimensions,''
Phys. Lett. B {\bf 142}, 173 (1984).

\end{thebibliography}
\end{document}